\documentstyle[12pt]{article}
\begin{document}

\def\r{\rightarrow}
\def\G{\Gamma}
\def\f{\phi}
\def\p{\partial}
\def\ii{\'{\char'20}}
\def\beq{\begin{equation}}
\def\brr{\begin{array}}
\def\err{\end{array}}
\def\eeq{\end{equation}}
\def\bea{\begin{eqnarray}}
\def\eea{\end{eqnarray}}
\def\bs{\bigskip}
\def\tr{\mbox{Tr}\, }
\def\ni{\noindent}
\def\wt{\widetilde}
\def\wh{\widehat}
\def\ul{\underline}
\def\nn{\nonumber}
\def\ms{\medskip}
\def\sp{\mbox{Sp}}
\def\re{\mbox{Re}\, }
\def\txs{\textstyle}
\def\dsp{\displaystyle}

\begin{titlepage}

\title{\begin{flushright}
{\normalsize UB-ECM-PF 94/4 }
\end{flushright}
\vspace{2cm}
{\Large \bf Possible evidences of Kaluza-Klein particles in a
scalar model with spherical compactification}}

\author{E. Elizalde \thanks{E-mail address:
eli@ebubecm1.bitnet, eli@zeta.ecm.ub.es} \\
Department E.C.M. and I.F.A.E., Faculty of Physics,
\\ University of  Barcelona, Av. Diagonal 647, 08028 Barcelona, \\
and Center for Advanced Studies, C.S.I.C., Cam\'{\i} de Santa B\`arbara,
\\
17300 Blanes, Catalonia, Spain \\
 and \\
\and Yu. Kubyshin \thanks{On leave of absence from Nuclear
Physics Institute,
Moscow State University, 117234 Moscow, Russia.}
\thanks{E-mail address: kubyshin@ebubecm1.bitnet} \\
Department E.C.M., Faculty of Physics, University of
Barcelona, \\
Av. Diagonal 647, 08028 Barcelona, Catalonia, Spain}

\date{February 14th, 1994}

\maketitle

\begin{abstract}

Possible experimental manifestations of the contribution of heavy
Kaluza-Klein particles, within a simple scalar model in six
dimensions
with spherical compactification, are studied. The approach is
based
on the assumption that the inverse radius $L^{-1}$ of the
space
of extra dimensions
is of the order of the scale of the supersymmetry breaking
$M_{SUSY}
\sim 1 \div 10$ TeV. The total cross section of the scattering of
two light particles is calculated to one loop order and the
effect
of the Kaluza-Klein tower is shown to be noticeable for
energies
$\sqrt{s} \geq 1.4 L^{-1}$.

\end{abstract}
\end{titlepage}


\section{Introduction}

Many of the modern approaches extending the Standard Model
include the hypothesis
of multidimensionality of the space-time (e.g. Kaluza-Klein type
theories,
supergravity, superstring theory; see for instance \cite{Duff} and
\cite{green-87},
and references therein). The extra dimensions are supposed to be
compactified, i.e. "curled up" to a compact manifold of a
characteristic scale $L$.

Various aspects of multidimensional models of gravity and
particle interactions at the classical level have been
intensively studied (see, for example, \cite{KK-review1},
\cite{KK-review2} for reviews). Some issues related to
quantum features of Kaluza-Klein theories were considered
in \cite{KK-quantum} - \cite{1-loop}. One of the interesting
problems is the search and calculation of characteristic
effects related to multidimensional nature of Kaluza-Klein
theories. In this paper we consider one of the effects of
this kind, which is essentially quantum.

In many models the scale of compactification $L$ is assumed
to be (or appears to be) of the order of the inverse Planck
mass $M_{Pl}^{-1}$
(see, for example, \cite{compact} and the reviews
\cite{KK-review2}). In this case additional dimensions could
reveal themselves only as
peculiar gravitational effects or at an early stage of the
evolution of the Universe. On the other hand,
there are some arguments in favour of a larger
compactification scale. One of them comes from  Kaluza-Klein
cosmology and stems from the fact that the density of heavy
Kaluza-Klein particles cannot be too large, in order not to
exceed the critical density of the Universe. Estimates
obtained in ref. \cite{kolb} give the bound $L^{-1} < 10^{6}$
GeV.

Other arguments are related to results of papers
\cite{kapl-88} and
\cite{ant-90}. In many of the above mentioned theories there
is usually a
much lower supersymmetry breaking scale $M_{SUSY}$. This scale
can be
naturally related to the compactification scale since, as it is
known,
supersymmetry in principle lowers under compactification
of a part of the
space-time dimensions \cite{Duff}.
Also, as it was shown in ref. \cite{ant-90} relating these two scales
may lead to cancellation of unwanted threshold corrections in
superstring
theories for certain compactifications. In its turn multidimensional
models, because of their non-renormalizability, apparently should be
considered as some kind of low energy effective theory stemming from
more fundamental theory, like supersting theory. This gives
additional motivations for considering Kaluza-Klein models with
$L^{-1} \sim M_{SUSY}$.

No natural mechanism providing compactification of the
space of extra dimensions with such scale is known
so far. In the present paper -- having the above
mentioned arguments in mind -- we would like to study physical
consequences
in a multidimensional model assuming that a compactification
of this kind is indeed possible. Our aim is to find an effect
which (in principle) could be measured experimentally
 and that, because of the hypothesis made that
$L^{-1}$ is of the order of a few TeV, could actually be
observed at future experiments in
supercolliders. This could yield decisive evidence about the
validity of the Kaluza-Klein hypothesis within a given model.
In performing such analysis, we
must start by understanding what kind of effects can be observed
and how
they depend on the topology of the space of extra dimensions
in general, rather than on the phenomenological details of the
given model. That is why
we will here restrict ourselves to a simple $\phi ^{4}$ scalar
model with two extra dimensions.

As is well known, by doing mode expansion, a multidimensional
model on the space-time $M^{4} \times K$ (where $K$ is a
compact manifold)
can be represented as an effective theory on $M^{4}$ with an
infinite set of particles, which is often referred to as the
Kaluza-Klein tower of
particles or modes. The spectrum of the four-dimensional
theory depends on the topology and geometry of $K$. The
sector of the lowest state (of the zero mode) describes light
particles (in the sense that their masses do not depend on
$L^{-1}$) and coincides with the
dimensionally reduced theory. Higher modes correspond to heavy
particles with masses $\sim L^{-1}$ called {\it pyrgons}
(from the Greek $\pi \upsilon \rho
\gamma os$, for ladder). It is the contribution of
pyrgons to physical quantities that might give evidence about
the existence of extra dimensions.

Because of its multidimensional character, the complete theory
is non-renormalizable. As we have mentioned before, this can
be understood as not being a basic difficulty if the theory is a
low-energy limit of
some more fundamental one. Whereas in a complete finite quantum theory
renormalization counterterms of the low energy limit
could be calculated, in the effective non-renormalizable
theory, we are discussing here, the corresponding coupling
constants must be considered as {\it phenomenological} ones.
Fortunately, the contribution of these counterterms to the finite
part of the amplitude is of  order
$(s/\Lambda ^2)^n$ \cite{green-87}, where $\sqrt{s}$ is the
energy of the
colliding particles, $\Lambda $ can be regarded as the
characteristic
scale of the complete theory ($\Lambda \sim M_{Pl}$ in the
case of
superstrings), and $n \geq 1$. Thus, when $\sqrt{s} \leq M_C
\ll\Lambda $ these contributions can be neglected.

As  was demonstrated in refs. \cite{decoup1}, \cite{decoup2},
in spite of the infinite number of fields in the theory on
$M^{4}$ ---and
related to this non-renormalizability--- the contributions of
the heavy particles to the renormalized quantities decouple
when the energy of the process is small enough so that $sL \ll 1$.
This ensures that the low-energy limit of a Kaluza-Klein
model is just the dimensionally
reduced theory including zero modes only. When energies are
comparable to $L^{-1}$, the contributions of the heavy modes
are not negligible. One may expect that, because of the
infinite number of
states in the Kaluza-Klein tower, some accumulation of
contributions
of heavy modes takes place, which leads to a noticeable effect
even
for energies $\sqrt{s} < 2 M_{1}$ ($M_{1}$ is the mass of the
first heavy mode), when direct production of the
excited states is still not possible. Of course, because of
the decoupling, actually only a few of the lowest states (with
masses up to
about $10 L^{-1}$ ) will contribute essentially at these
energies.

The purpose of our paper is to demonstrate that a noticeable
effect does indeed appear. With this aim, we will choose a
simple scalar
$\phi ^{4}$-model on the six-dimensional space-time
$M^{4} \times S^{2}$. The space of extra dimensions is the
two-dimensional sphere $S^{2}$ of radius $L$ and with an
$SO(3)$-invariant metric on it. Study of quantum effects on
the spheres (mainly calculation of the Casimir effect and the
effective potential) can be found, for example, in \cite{Casimir},
\cite{V-sphere}.
In this paper we calculate
the total cross section for the \ (2 {\em light particles})
$\longrightarrow$ \ (2 {\em light particles}) \ scattering process.
Pyrgons do not contribute at
tree level in this model. Actually, the discussion given in
Sect. 2  shows
that this statement is rather general. Thus, the 1-loop
correction is important for having the effect due to presence
of heavy Kaluza-Klein modes. We will calculate the 1-loop
contribution and analyze it in a wide range of energies of the
scattering particles. A similar problem but for the
torus compactification (i.e. when
the space of extra dimensions $K$ is the two-dimensional torus
$T^{2}$) was considered in ref. \cite{DKPC}. We will use the
results of
that paper to compare with the results obtained in the present one,
in order to understand to which extent the cross section may
depend on
the topology of the space of extra dimensions. We believe that
our results are rather generic, and that comparable effects
could be detected in
more realistic theories constructed in the framework of the
Kaluza-Klein approach. We would like to mention that some results
on calculation of 1-loop Feynman diagrams on the space-time
$M^{4} \times T^{N}$ can be found in \cite{1-loop}.

An alternative attitude was taken in ref. \cite{Kos-90}.
There the authors
used the results of high energy
experiments to get an upper bound on the size $L$ of the space
of extra
dimensions, assuming that even the first heavy Kaluza-Klein
mode is not observed
experimentally. Our philosophy in the present paper is
different: we
assume that $L^{-1}\sim M_{SUSY}\sim 1\div 10$ TeV and look
for possible
experimental evidence of heavy Kaluza-Klein modes. It is clear
that
our results can be also used for obtaining bounds on $L$,
provided
that no experimental evidence of extra dimensions is seen in
future experiments for this range of energies.

The paper also contains a section of more mathematical character, in
which some zeta function regularization techniques \cite{hawk} are
described that provide a rather elegant way of treating sums over
Kaluza-Klein modes which appear in the
theory (for a review of these methods see \cite{eorbz}).
Though there is some literature on the technique of performing
calculations on the spheres (see, for example, \cite{Casimir},
\cite{V-sphere}, \cite{1-loop} and references therein), we think
that this part of our work is interesting in its own right,
since some of the formulas that appear there are brand new
and provide a non-trivial, alternative way of to deal with
spherical compactification.
Some general results on calculations on curved space-time can
be found in the review \cite{camporesi}

The paper is organized as follows. In Sect. 2 we describe the
model,
choose the renormalization condition and discuss the general
structure of the 1-loop results. In Sect. 3,
some useful asymptotic expansions of the
corresponding (regularizing) zeta function are derived.
Methods for the
numerical evaluation of the 1-loop contribution are
devised and the
amplitude and cross section for a wide range of energies is
 calculated.
Sect. 4 contains
the analysis of the total cross
section in the multidimensional model, and its comparison with the cross
section for theories with a finite number of
heavy particles, as well as with
the cross section obtained for the torus compactification.
Concluding remarks are presented in Sect. 5.

\section{Description of the model, mode expansion and
renormalization}

Let us consider a one component scalar field on the
$(4+2)$-dimensional
manifold $E=M^{4}\times S^{2}$, where $M^{4}$ is Minkowski
space-time and
$S^{2}$ is a two-dimensional sphere of radius $L$. In spite of
its
simplicity this model captures some
interesting features of both the classical and quantum properties
of
multidimensional theories. The action is given by
\begin{equation}
S= \int_{E} d^{4}x d\Omega \left[\frac{1}{2}(\frac{\p \f
(x,\theta)}{\p x^{\mu}})^2+\frac{1}{2} g^{ij} \frac{\p \f (x,\theta)}
{\p \theta^{i} } \frac{\p \f (x,\theta) }{\p \theta^{j} } -
 \frac{1}{2} m_{0}^{2} \f ^{2}(x,\theta) -
\frac{\hat{\lambda}}{4!} \f ^{4} (x,\theta) \right],
\label{eq:action0}
\end{equation}
where $x^{\mu}, \mu = 0,1,2,3,$ are the coordinates on
$M^{4}$, $\theta^{1}$ and $\theta^{2}$ are the
standard angle coordinates on $S^{2}$, $0
< \theta^{1} < \pi$, $0 < \theta^{2} < 2 \pi$, and $d \Omega$
is the integration measure on the sphere. The metric $g_{ij}$ is
the standard
$SO(3)$-invariant metric on the two-dimensional sphere:
\[ ds^{2} = g_{ij} d\theta^{i} d\theta^{j} = L^{2}
[(d\theta^{1})^{2}
+ \sin ^{2} \theta^{1} (d\theta ^{2})^{2}].  \]
To re-interpret this model in four-dimensional terms we
make an expansion of the field $\phi (x,\theta)$,
\begin{equation}
 \phi (x,\theta) = \sum_{lm} \phi _{lm} (x) Y_{lm} (\theta),
\label{eq:expansion}
\end{equation}
where $l = 0,1,2, \ldots$, $m=-l,-l+1, \ldots, l-1,l$ and
$Y_{lm}(\theta)$ are the
eigenfunctions of the Laplace operator on the internal space,
i.e. spherical harmonics, satisfying
\begin{eqnarray}
   \Delta Y_{lm} & = & - \frac{l(l+1)}{L^{2}}  Y_{lm},
\label{eq:laplace} \\
\int d \Omega Y_{lm}^{*} Y_{l'm'} & = &\delta _{l,l} \delta
_{m,m'}.
\label{eq:orthonorm}
\end{eqnarray}
Substituting this expansion
into the action and integrating over $\theta$, one obtains
\begin{eqnarray}
   S & = & \int_{M^{4}} d^{4} x \left\{ \frac{1}{2} ( \frac{\p
\f_{0} (x)} {\p x^{\mu}} )^{2}
- \frac{1}{2} m_{0}^{2} \f_{0}^{2}(x) -
\frac{\lambda_{1}}{4!} \f _{0}^{4} (x) \right.  \nonumber  \\
     &  & + \sum_{l>0} \sum_{m} \left[ \frac{\p \f_{lm}^{*} (x)}{\p
x^{\mu}}
\frac{\p \f_{lm} (x)}{\p x_{\mu}} - M_{l}^{2} \f _{lm}^{*} (x)
\f _{lm} (x) \right]  \nonumber  \\
     &  & - \left. \frac{\lambda_{1}}{2}
\f _{0}^{2} (x) \sum_{l > 0} \sum_{m} \f_{lm}^{*}(x)
\f_{lm}(x) \right\} - S'_{int},
\label{eq:action1}
\end{eqnarray}
where the four-dimensional coupling constant $\lambda_{1}$ is
related to the
multidimensional one $\hat{\lambda}$ by
$ \lambda_{1} = \hat{\lambda} / {\mbox volume} (S^{2})$.
In (\ref{eq:action1})
$S_{int}'$ includes all terms containing third and fourth
powers of
$\f_{lm}$ with $l > 0$. We see that (\ref{eq:action1})
includes one
real scalar field $\f_{0} \equiv \f_{00}(x)$ describing a
light particle of mass $m_{0}$, and an infinite set
(``tower") of massive complex fields $\f_{lm}(x)$ corresponding
to heavy particles, or pyrgons, of masses
given by
\begin{equation}
  M_{l}^{2} = m_{0}^{2} + l(l+1) M ,  \label{eq:mass}
\end{equation}
where $M = L^{-1}$ is the compactification scale.

Let us consider the 4-point Green function $\G ^{(\infty)}$
with external legs corresponding to the light particles
$\f _{0}$. The index
$(\infty)$ indicates that the whole Kaluza-Klein tower of
modes
is taken into account. Diagrams
which contribute to this function in the tree and one-loop
approximation are presented in Figs. 1 and
 2. Terms included in $S'_{int}$
in (\ref{eq:action1}) are not relevant for our computations.

Let us first analyze the tree level contribution. The first
diagram in Fig. 1 is exactly the same as in the case of
the dimensionally reduced theory whose action
is given by the first line in eq.
(\ref{eq:action1}), the second diagram appeares due to
extra divergences in the theory and is discussed below.
Heavy modes do not contribute at this level. This
property is rather general and is valid for all processes of
the type \ ($n$ {\em light particles}) \ $\longrightarrow$
\ ($k$ {\em light particles}) \ at least for
all scalar multidimensional theories with polynomial
interactions.
The reason for this is rather simple. Suppose that we have a
theory
on $M^{4} \times K$, where $K$ is a compact space, with a
polynomial interaction. The analysis of tree graphs
shows that in order to obtain a contribution
of heavy modes at the tree level, one needs to have at least
one vertex at which $q$ light modes $\f _{0}(x)$ (with
$q < n+k $ for $(n+k) > 2$) interact with one heavy mode
$\f _{N}(x)$, where the generalized index $N$ corresponds to
a non-zero
eigenvalue of the Laplace operator on $K$. After substituting
the expansion of the multidimensional field over the
eigenfunctions of
this operator (analogous to (\ref{eq:expansion}))
into the original multidimensional action,
the interaction term
\[  \int_{M^{4}} d^{4}x \f _{N}(x) (\f _{0}(x))^{q},     \]
corresponding to this vertex, will always appear multiplied by
a factor
\[   \int_{K} d\Omega Y_{N}(\theta) \underbrace{Y_{0} \ldots
Y_{0}}_
{q \ \mbox{times}}. \]
But since the eigenfunction $Y_{0}$ corresponding to the zero
eigenvalue of the Laplace operator on the compact manifold is
a constant,\footnote{The authors thank C. Nash for useful
discussions on this issue.} the above integral will always  be
zero, due to the orthonormality condition. This implies, for
example, that the process of decay of a heavy mode with number
$N$ into $q$ zero modes is forbidden for the class of models
under consideration.
In the case $K = S^{2}$ the feature discussed here is a
manifestation of the conservation of angular momentum.

Let us analyze now the 1-loop correction in our model.
It is easy to check that, owing to the infinite sum of
diagrams (see Fig. 2), the Green function to one loop order is
quadratically divergent. This is certainly a reflection of the fact
that the original theory is actually six-dimensional and, therefore,
non-renormalizable.
Thus, the divergencies cannot be removed by renormalization of
the coupling constant in (\ref{eq:action1}) alone. We must
also add a
counterterm $\lambda_{2B}\f ^{2}(x,y) \Box _{(4+d)} \f^{2}(x,y)$,
 where $\Box _{(4+d)}$ is the d'Alambertian on $E$ and
$\lambda_{2B}$ has mass dimension two. The second diagram in Fig. 1
corresponds to the contribution of this counterterm. Of
course, for calculation of other Green functions, or
higher-order loop
corrections, other types of counterterms are necessary, but we
are not going to discuss them here. Hence, the Lagrangian  we
will use for our investigation is
\begin{eqnarray}
{\cal L} & = & \frac{1}{2} ( \frac{\p \f _{0}(x)} {\p
x} )^{2} + \frac{m_{0}^{2}}{2} \f ^{2}_{0}(x)  \nonumber \\
  & + & \sum_{l > 0} \sum_{m} \left[ \frac{\p \f_{lm}^{*} (x)}{\p
x^{\mu}}
\frac{\p \f_{lm} (x)}{\p x_{\mu}} - M_{l}^{2} \f _{lm}^{*} (x)
\f _{lm}
(x) \right]        \nonumber \\
  & - & \frac{\lambda_{1B}}{4!} \f _{0}^{4} (x) -
\frac{\lambda_{1B}}{2}
\f _{0}^{2} (x) \sum_{l > 0} \sum _{m} \f_{lm}^{*}(x)
\f_{lm}(x) -
\frac{\lambda_{2B}}{4!} \f _{0}^{2}(x) \Box \f _{0}^{2}(x),
\label{eq:action2}
\end{eqnarray}
where $\lambda _{1B}$ and $\lambda_{2B}$ are bare coupling
constants. To regularize the four-dimensional integrals
corresponding to the 1-loop diagrams of Fig. 2, we will employ
dimensional regularization, which is performed, as usual, by
making analitical continuation of the integrals to $(4-2\epsilon)$
dimensions. $\kappa$ will be a mass
scale set up by the regularization procedure. The sums over $l$ will
be regularized by means of the zeta function technique \cite{hawk}.

Let us now specify the renormalization scheme. It is known
that the results of the calculation of physical
quantities at finite orders of perturbation theory depend
on the renormalization scheme. Since our goal is
basically to study the difference between the contributions
corresponding to the complete
Kaluza-Klein tower of particles and to the light particle
only, we should
choose our scheme such that the results are minimally affected
by the
renormalization procedure. A reasonable way to do this
is to impose the condition that the physical amplitude
for the \ (2 {\em light particles}) \ $\longrightarrow$ \ (2 {\em
light particles}) \ scattering process
of the complete theory and the amplitude of the same process
in the four-dimensional theory with the zero mode only (i.e. the
theory
given by the Lagrangian (\ref{eq:action2}) with all non-zero
modes
and the last term being omitted) must coincide at some
normalization
(subtraction) point corresponding to low energies. As
subtraction point we will choose the following point in
the space of invariant variables built up out of the external
four-momenta  $p_{i}$ $(i=1,2,3,4)$  of the scattering
particles:
\begin{eqnarray}
& & p_{1}^{2}  =  p_{2}^{2} =   p_{3}^{2} =  p_{4}^{2} =
m_{0}^{2},
\nonumber \\
 & & p_{12}^{2} = s = \mu_{s}^{2}, \; \;
     p_{13}^{2} = t = \mu_{t}^{2}, \; \;
     p_{14}^{2} = u = \mu_{u}^{2},
 \label{eq:point}
\end{eqnarray}
where $p_{1j}^{2} = (p_{1} + p_{j})^{2}$, $(j=2,3,4)$, and
$s$, $t$ and $u$ are the Mandelstam variables.
Since the subtraction point is located on the mass shell, it
satisfies the standard relation
$\mu_{s}^{2} + \mu_{t}^{2} + \mu_{u}^{2} = 4 m_{0}^{2}$. The
renormalization prescription formulated above can be written
as
\begin{equation}
 \left. \G ^{(\infty)} (p_{1j}^{2}; m_{0}, M, \lambda_{1B},
\lambda_{2B}, \epsilon) \right|_{s.p.} =
  \left. \G ^{(0)} (p_{1j}^{2}; m_{0}, \lambda_{1B}',
\epsilon) \right|_{s.p.}  =  g \kappa ^{2\epsilon},
\label{eq:renorm1}
\end{equation}
\begin{equation}
 \left. \left[ \frac{\p}{\p p_{12}^{2}} + \frac{\p }{\p p_{13}^{2}} +
 \frac{\p }{\p p_{14}^{2}}\right]
\G^{(\infty)} \right|_{s.p.} =
 \left. \left[ \frac{\p}{\p p_{12}^{2}} + \frac{\p }{\p p_{13}^{2}} +
 \frac{\p }{\p p_{14}^{2}}\right]
\G^{(0)} \right|_{s.p.} + \frac{\lambda_{2}}{4} \kappa ^{-2 +
2\epsilon}.     \label{eq:renorm2}
\end{equation}
Here $\G ^{(0)}$ is the four-point Green function of the
four-dimensional theory with the zero mode field only (i.e.,
the dimensionally reduced theory), $\lambda_{1B}'$ being its
bare coupling
constant. In the first line we have written down
the dependence of the Green functions on the momentum
arguments and
parameters of the theory explicitly, and we have taken into
account that to one-loop order they depend on $p_{12}^{2}$,
$p_{13}^{2}$, and $p_{14}^{2}$ only. The label $s.p.$ means that the
corresponding quantities are taken at the subtraction point
(\ref{eq:point}). $g$
and $\lambda_{2}$ are renormalized coupling constants. The
last one is included for the sake of generality only, and we
will see that our result does not depend on it.

For the usual $\lambda \f_{0}^{4}$-theory in four dimensions,
the renormalization condition given by (\ref{eq:renorm1})
alone is sufficient, whereas both conditions
(\ref{eq:renorm1}) and (\ref{eq:renorm2}) are necessary
for subtracting the ultraviolet
divergences in the theory (\ref{eq:action2}), because of the
presence of additional
divergences due to its multidimensional character.

To one-loop order, the Green functions of the complete theory and of
the theory with only zero mode, are given by
\begin{eqnarray}
\G ^{(\infty)} (p_{1j}^{2}; m_{0}, M, \lambda_{1B}, \lambda_{2B},
\epsilon) & = &\lambda_{1B} +
\lambda_{2B} \frac{p_{12}^{2}+p_{13}^{2}+p_{14}^{2}}{12}
\nonumber
\\
 & + & \lambda_{1B}^{2} [ K_{0}(p_{1j}^{2}; m_{0},\epsilon) +
       \Delta K (p_{1j}^{2}; m_{0},M,\epsilon) ], \label{eq:1loop}
\\
 \G ^{(0)} (p_{1j}^{2}; m_{0}, \lambda_{1B}',\epsilon) & = &
   \lambda_{1B}' + \lambda_{1B}'^{2} K_{0}(p_{1j}^{2}; m_{0}, \epsilon).
\nonumber
\end{eqnarray}
Here
\begin{equation}
K_{0}(p_{1j}^{2};m_{0},\epsilon) \equiv K_{00}(p_{1j}^{2};m_{0}^{2},
\epsilon), \; \; \;  \Delta K (p_{1j}^{2};m_{0},M,\epsilon)=
\sum_{l=1}^{\infty} \sum_{m=-l}^{l} K_{lm}(p_{1j}^{2};M_{l}^{2},\epsilon)
\label{eq:Kexpansion} ,
\end{equation}
and $K_{lm}$ is the contribution of the mode $\f _{lm}$
with  mass $M_{l} = \sqrt{m_{0}^{2} + M^{2} l(l+1)}$ (see eq.
(\ref{eq:mass})) to
the one-loop diagram of two light particles scattering
\begin{equation}
K_{lm} (p_{1j}^{2}; M_{l}^{2},\epsilon) =
\frac{-i}{32 \pi^{4}} \frac{1}{M_{l}^{2\epsilon}}
\left[ I(\frac{p_{12}^{2}}{M_{l}^{2}}, \epsilon) +
I(\frac{p_{13}^{2}}{M_{l}^{2}}, \epsilon) +
I(\frac{p_{14}^{2}}{M_{l}^{2}}, \epsilon)\right].
    \label{eq:Kdefinition}
\end{equation}
$K_{0}$ corresponds to the first diagram in Fig. 1, and $\Delta K$
 to the contribution of the sum term in Fig. 2. Here we assume
that $\lambda_{2B} \sim \lambda_{1B}^{2}$, so that
the one loop diagrams proportional to
$\lambda_{1B} \lambda_{2B}$ or $\lambda_{2B}^{2}$
can be neglected. It can be shown that this
hypothesis is consistent (see ref. \cite{decoup1}).
Notice that if we did not make this assumption,
control on the proliferation of divergences would become very hard.
The function $I$ in the formula above is the standard one-loop
integral
\begin{eqnarray}
I(\frac{p^{2}}{M^{2}},\epsilon )
   & = & M^{2\epsilon} \int d^{4-2\epsilon} q \
         \frac{1}{(q^{2}+M^{2})((q-p)^{2}+M^{2})}
\nonumber \\
   & = & i \pi^{2-\epsilon} \Gamma (\epsilon)
         M^{2\epsilon} \int_{0}^{1} dx \ \frac{1}
         {[M^{2} - p^{2} x(1-x)]^{\epsilon}}.
\label{eq:integral}
\end{eqnarray}
Let us also introduce the sum of the one-loop integrals over
all Kaluza-Klein modes
\begin{equation}
  \Delta I (\frac{p^{2}}{M^{2}},\frac{m_{0}}{M},\epsilon) =
\sum_{l=1}^{\infty}
  \sum_{m=-l}^{l} (\frac{M^{2}}{M_{l}^{2}})^{\epsilon}
  I(\frac{p^{2}}{M_{l}^{2}},\epsilon ),
\label{eq:int-sum}
\end{equation}
so that
\[ \Delta K (p_{1j}^{2}; m_{0},M,\epsilon) = \frac{-i}{32 \pi ^{4}}
 \frac{1}{M^{2 \epsilon}}
 \left[ \Delta I (\frac{p_{12}^{2}}{M^{2}},\frac{m_{0}}{M},\epsilon) +
   \Delta I (\frac{p_{13}^{2}}{M^{2}},\frac{m_{0}}{M},\epsilon) +
   \Delta I (\frac{p_{14}^{2}}{M^{2}},\frac{m_{0}}{M},\epsilon) \right]
. \]
Performing the renormalization, we obtain the following
expression
for the renormalized four-point Green function
\begin{eqnarray}
  & \G ^{(\infty)}_{R}&
  \left(\frac{p_{1j}^{2}}{\mu_{j}^{2}}; \frac{ \mu_{j}^{2}}{M ^{2}};
  \frac{m_{0}}{M}, \frac{\kappa}{M}, g, \lambda_{2}\right) =
  \lim _{\epsilon \rightarrow 0} \kappa ^{-2 \epsilon}
  \G ^{(\infty)} \left(p_{1j}^{2}; m_{0}, M, \lambda_{1B} (g,
\lambda_{2}\right),
  \lambda_{2B}(g, \lambda_{2}), \epsilon)  \nonumber \\
  & = &  \lim _{\epsilon \rightarrow 0} \left\{ g + \lambda_{2}
  \frac{p_{12}^{2}+p_{13}^{2}+p_{14}^{2}-\mu_{s}^{2}-\mu_{t}^{2}-\mu_{u}^{2}}
  {12 \kappa^{2}}   \right.
 \label{eq:gfren1}
    \\
  & + & g^{2} \kappa ^{2 \epsilon} \left[ K_{0}(p_{1j}^{2};
m_{0},\epsilon) -
  K_{0}(\mu_{j}^{2}; m_{0},\epsilon) \right.
  \nonumber \\
  & + & \Delta K (p_{1j}^{2}; m_{0},M,\epsilon) -
  \Delta K (\mu_{j}^{2}; m_{0},M,\epsilon)
  \nonumber \\
  & - & \left. \left. \left.
\frac{p_{12}^{2}+p_{13}^{2}+p_{14}^{2}-\mu_{s}^{2}-
  \mu_{t}^{2}-\mu_{u}^{2}}{3} ( \frac{\p }{\p p_{12}^{2}} +
  \frac{\p }{\p p_{13}^{2}} + \frac{\p }{\p p_{14}^{2}} )
       \Delta K (p_{1j}^{2}; m_{0},M,\epsilon) \right|_{s.p.} \right]
\right\}, \nonumber
\end{eqnarray}
where we denote $\mu_{2}^{2} = \mu_{s}^{2}$, $\mu_{3}^{2} = \mu_{t}^{2}$
and $\mu_{4}^{2} = \mu_{u}^{2}$.
The r.h.s. of this expression is regular in $\epsilon$, and
after calculating
the integrals and the sums over $l$ and $m$, we take the
limit $\epsilon \rightarrow 0$. The above expression is rather
general and valid for an arbitrary subtraction
point. It simplifies if the relation $\mu_{s}^{2} +
\mu_{t}^{2} + \mu_{u}^{2}
= 4 m_{0}^{2}$, fulfilled by the subtraction point
(\ref{eq:point}), is taken
into account. Before doing this, let us discuss the structure
of the
renormalized
Green function for a subtraction point when all subtraction
scales are equal:
$\mu_{s}^{2}=\mu_{t}^{2}=\mu_{u}^{2}=\mu^{2}$. Then eq.
(\ref{eq:gfren1}) can
be written as
\begin{eqnarray}
  && \G ^{(\infty)}_{R} (\frac{p_{1j}^{2}}{\mu^{2}};
  \frac{ \mu^{2}}{M ^{2}};
  \frac{m_{0}}{M}, \frac{\kappa}{M}, g, \lambda_{2})
    \nonumber \\
  &  & \ \ \ = g + \lambda_{2} \frac{p_{12}^{2}+p_{13}^{2}+p_{14}^{2}
  -3 \mu^{2}} {12 \kappa^{2}} \nonumber \\
  &  & \ \ \ +  g^{2} \lim _{\epsilon \rightarrow 0} \kappa ^{2 \epsilon}
  \left\{ K_{0}(p_{1j}^{2}; m_{0},\epsilon) -
  K_{0}(\mu^{2},\mu^{2},\mu^{2}; m_{0},\epsilon)
  - \frac{i}{32 \pi ^{4}} \frac{1}{M^{2\epsilon}} \left[
\left(
  \Delta I(\frac{p_{12}^{2}}{M^{2}},\frac{m_{0}}{M},\epsilon)
  \right. \right. \right.
                                \label{eq:gfrens} \\
  &  &  \left. \left. - \Delta I (\frac{\mu^{2}}{M^{2}},
  \frac{m_{0}}{M},\epsilon) - (p_{12}^{2} - \mu^{2})
  \frac{\p}{\p p_{12}^{2}}
  \Delta I (\frac{p_{12}^{2}}{M^{2}},\frac{m_{0}}{M},\epsilon)
  \right|_{p_{12}^{2}=\mu^{2}} \right) \nonumber \\
  &  & \ \ \ + \left. \left. ( p_{12}^{2} \longrightarrow p_{13}^{2} )
  + ( p_{12}^{2} \longrightarrow p_{14}^{2}) \right] \right\}.
\nonumber \end{eqnarray}
As it was mentioned above, the contribution $\Delta I$ of the
heavy modes contains quadratic
divergences in momenta (in the framework of the dimensional
regularization
used here this means that it contains singular terms $\sim
1/\epsilon$
and $\sim p^{2}/ \epsilon$). The renormalization prescription
amounts to a
subtraction of the first two terms of the Taylor expansion of
this contribution at the point $\mu^{2}$, which is sufficient
to remove the
divergences. The contribution $K_{0}$ of the zero mode sector
(the
dimensionally reduced theory) diverges only logarithmically
and subtraction
of the first term of the Taylor expansion, imposed by the
renormalization
prescription (\ref{eq:renorm1}), is sufficient to make it
finite
in the limit $\epsilon \longrightarrow 0$.

Finally, let us write down the expression for the four-point
Green
function of the complete theory (i.e. with all the
Kaluza-Klein modes)
renormalized according to the conditions (\ref{eq:renorm1}) and
(\ref{eq:renorm2})
at the subtraction point (\ref{eq:point}), and taken at a
momentum point which lies
on the mass shell of the light particle
\begin{eqnarray}
&&   \G ^{(\infty)}_{R}
   (\frac{s}{\mu_{s}^{2}},\frac{t}{\mu_{t}^{2}},\frac{u}{\mu_{u}^{2}};
   \frac{ \mu_{s}^{2}}{M ^{2}},\frac{ \mu_{t}^{2}}{M ^{2}},
   \frac{ \mu_{u}^{2}}{M ^{2}};
  \frac{m_{0}}{M}, g)
    \nonumber   \\
  &   & \ \ \ = g + g^{2} \lim _{\epsilon \rightarrow 0}
  \kappa ^{2 \epsilon} [ K_{0}(s,t,u;m_{0},\epsilon) -
  K_{0}(\mu_{s}^{2},\mu_{t}^{2},\mu_{u}^{2}; m_{0},\epsilon)
\nonumber \\
  &  &  \ \ \ +
  \Delta K (s,t,u; m_{0},M,\epsilon) -
  \Delta K (\mu_{s}^{2},\mu_{t}^{2},\mu_{u}^{2};m_{0},M,\epsilon) ].
  \label{eq:gfren2}
\end{eqnarray}
The variables $s$, $t$ and $u$ are not independent, since
they satisfy the
well known Mandelstam relation $s+t+u=4m_{0}^{2}$.

The formula
above is rather remarkable. It turns out that on mass shell, due
to cancellations between the
$s$-, $t$- and $u$-channels, the
contribution proportional
to $\lambda_{2}$ and the terms containing derivatives of the
one-loop
integrals vanish. Thus, heavy Kaluza-Klein modes contribute to
the renormalized Green function on the mass shell in exactly the same
way as
the light particle in the dimensionally reduced theory does.
Indeed, it can be
easily checked that the additional non-renormalized
divergences
arising from the infinite summation in $\Delta K$ cancel among
themselves when the three scattering channels are summed up
together.

\section{ Calculation of the one-loop contribution}

In this section we will analyze the one-loop
contribution of the heavy Kaluza-Klein modes, develop methods
for its numerical evaluation, and present results for the
amplitude of \ (2 {\em light particles}) \  $\longrightarrow$ \ (2
{\em light particles}) \  scattering.

Starting point of the analysis is the expression
\beq
 \Delta I(\frac{p^2}{M^2},\frac{m_{0}}{M},\epsilon) =
 i \pi^{2-\epsilon} \Gamma (\epsilon ) \int_0^1 dx \,
{\sum_{l,m}}' \left( \frac{M^{2}}{M_{l}^2} \right)^{\epsilon}
\left[ 1- \frac{p^2x(1-x)}{M_{l}^2} \right]^{-\epsilon},
\label{i11}
\eeq
(see (\ref{eq:integral}) and (\ref{eq:int-sum})),
where the prime means that the term for $l=0$ is absent from the summatory.
One could think of considering two different limits
corresponding to two
possible expansions of the binomial (low and high momentum
expansions, in principle):
 \bea
(i) && (1-y)^{-\epsilon} = \sum_{k=0}^\infty  \frac{\Gamma
(k+\epsilon)}{k!\, \Gamma (\epsilon)} \, y^k, \ \ \ \ \ \ \ \
y\equiv
\frac{p^2x(1-x)}{M_{l,m}^2}, \\
(ii) && (1-y)^{-\epsilon} =
(-y)^{-\epsilon}\left(1-y^{-1}\right)^{-\epsilon}
= (-1)^{-\epsilon} \sum_{k=0}^\infty  \frac{\Gamma
(k+\epsilon)}{k!\, \Gamma (\epsilon)} \, y^{-k-\epsilon}.
\label{hme}
\eea

There is no problem in doing the small-momentum expansion
(i), which is valid for $|y|<1$. In fact, since the maximum of
$x(1-x)$
when $0\leq x \leq 1$ is attained at $1/4$ (for $x=1/2$), this
formula is valid whenever $p^2 < 4M^2_{l,m}$. Nevertheless,
the ``high-momentum expansion"
(ii) is much more difficult to perform. Actually, it is {\it
not}
possible
to express its range of validity,  $|y|<1$, in terms of a
simple
inequality involving $p^2$ and $M^2_{l,m}$. As it stands, eq.
(\ref{hme}) is useless:
we must first integrate over $x$, in order to get rid of this
unwanted
dependence and  then  the formula yielding
the desired expansion for  $p^2 \geq 4M^2_{l,m}$ is
different,
acording to different ranges of variation of  $p^2$ in terms
of
$M^2_{l,m}$.

Taking into account eq. (\ref{eq:mass}), we will see that
the infinite sum over $l$  gives rise to a certain derivative of
a generalized, inhomogeneous
Epstein zeta-function, which we will manage to continue analytically
after some work.
On the other hand, in case (i) the $x$-integral yields just  beta
function factors.
\ms

\ni{\bf 3.1. Expansion for small momentum}.
\ms

The sums and integrals involved in the low-momentum expansion
of
eq. (\ref{i11}) can be performed in the following order:
\beq
 \Delta I(\frac{p^{2}}{M^{2}}, \frac{m_{0}}{M}, \epsilon)
 = i \pi^{2-\epsilon} \sum_{k=0}^\infty
\frac{\Gamma (k + \epsilon)}{k!} \, B(k+1,k+1) \left(
\frac{p^{2}}{M^2}
\right)^k S_{k+\epsilon}, \eeq
where we have defined
\beq
S_{k+\epsilon} \equiv {\sum_{l,m}}' \left(
\frac{M^2_{l,m}}{M^2}
\right)^{-k-\epsilon},
\eeq
$M^2 = (L^{-1})^{2}$ being the constant, leading mass which
invariably appears
in $M^2_{l}$ (see (\ref{eq:mass}) for the sphere
compactification),
and where we have used
\beq
\int_0^1 dx\, [x(1-x)]^s = B(s+1,s+1),
\eeq
$B(s,t)= \Gamma (s)\Gamma (t)/\Gamma (s+t)$ being Euler's beta
function.

In the particular case of the spherical compactification, this
sum reads
\beq
S_{k+\epsilon} = \sum_{l=1}^\infty (2l+1) \left[ l(l+1) +
\frac{m_{0}^2}{M^2} \right]^{-k-\epsilon} = 2  \sum_{l=1}^\infty
(l+1/2) \left[ (l+1/2)^2 +\left( \frac{m_{0}^2}{M^2}- \frac{1}{4}
\right) \right]^{-k-\epsilon},
\eeq
and can be written exactly as
\beq
S_{k+\epsilon}= \frac{1}{1-k-\epsilon} \,\left.
\frac{\partial}{\partial a}
F(s;a,b)\right|_{ \dsp s=k+\epsilon, a=\frac{1}{2}, b=
\frac{m^2}{M^2}- \frac{1}{4}},
\label{ske}
\eeq
where the function $F$ is defined to be
\beq
F(s; a,b) \equiv \sum_{l=1}^\infty [(l+a)^2+b]^{1-s},
\label{fm}
\eeq
and is related to the most simple example of ---what is called--- an
Epstein-Hurwitz zeta function. Some useful and
mathematically elegant expressions for
these functions have been obtained in refs. \cite{eli1} (see also
\cite{eorbz}).
In particular, use of Jacobi's theta function identity yields
\cite{eli1}
\bea
F(s;a,b) &=& \frac{b^{1-s}}{\Gamma (s-1)} \sum_{n=0}^\infty
\frac{(-1)^n\Gamma (n+s-1)}{n!} \, b^{-n} \zeta (-2n,a) +
\frac{\sqrt{\pi}}{2} \, b^{3/2-s} \, \frac{\Gamma
(s-3/2)}{\Gamma
(s-1)} \nn \\
&&+ \frac{2\pi^{s-1}}{\Gamma (s-1)} \,
b^{3/4-s/2}
\sum_{n=1}^\infty n^{s-3/2} \cos (2 \pi n a) \, K_{s-3/2} (2\pi n
\sqrt{b}), \label{eq:repres} \eea
which, in spite of the equality sign, should be understood as an
asymptotic expression, {\it not} as a convergent series expansion.
Later we will obtain explicitly the optimal cut of this series, which
has been numerically studied in detail in Ref. \cite{eeb3}. Taking now
into account eq. (\ref{ske}), we obtain
\beq S_{k+\epsilon}= \sum_{n=0}^\infty \frac{(-1)^{n-1}}{n!} \,
\frac{\Gamma (n+k+\epsilon -1)}{\Gamma (k+\epsilon)}
\,  b^{1-n-k-\epsilon} \, \wt{\zeta} (-2n,1/2),
\eeq
where
\beq
 \wt{\zeta} (s,a) \equiv  \frac{\partial}{\partial a} \zeta
(s,a),
\eeq
$ \zeta (s,a)$ being the Hurwitz (also called Riemann
generalized)
zeta function defined for $s > 1$ by the series
\beq
 \zeta (s,a) \equiv \sum_{l=0}^{\infty} \frac{1}{(l+a)^{s}}.
\eeq
Actually, the following simple relation holds
\beq
 \wt{\zeta} (s,a) = -s \zeta (s+1,a),
\label{wt1}
\eeq
and for the few first terms of $S_{k+\epsilon}$ (providing the
best
cut
of the asymptotic series), we obtain
\bea
S_{k+\epsilon}&=&   b^{1-k-\epsilon} \left[
\frac{1}{k+\epsilon -1}
+2b^{-1} \zeta(-1,1/2)- 2(k+\epsilon) b^{-2} \zeta(-3,1/2)
\right.
\nn
\\ && + \left. (k+\epsilon)(k+1+\epsilon) b^{-3}
\zeta(-5,1/2)-
\cdots
\right].
\eea
We have used the fact that the coefficient $-\wt{\zeta} (0, 1/2)=1$,
actually
\beq
\wt{\zeta} (0,a) = \frac{\partial}{\partial a} \zeta (0,a)
 = \frac{\partial}{\partial a} \left( \frac{1}{2} -a \right)
=-1.
\eeq
The same result is obtained from (\ref{wt1}) in the limit $s
\rightarrow 0$.

Putting everything together, we can write
\bea
 \Delta I(\frac{p^2}{M^2},\frac{m_{0}}{M},\epsilon) &=&
 i \pi^{2-\epsilon} \Gamma (\epsilon ) \left\{ \left[
\frac{b^{1-\epsilon}}{\epsilon -1} + 2 b^{-\epsilon} \zeta (-
1,1/2)- 2\epsilon b^{-1-\epsilon} \zeta (-3,1/2) \right.
\right.
\nn \\
&&+ \left. \epsilon
(1+\epsilon) b^{-2-\epsilon} \zeta (-5,1/2)- \frac{\epsilon
(1+\epsilon) (2+\epsilon)}{3} \, b^{-3-\epsilon} \zeta
(-7,1/2)+ \cdots
\right]  \nn  \\
  & + & \frac{p^2}{M^2} B(2,2)  \left[ b^{-\epsilon} + 2
\epsilon
b^{-1-\epsilon} \zeta (-1,1/2) - 2\epsilon (1+\epsilon) b^{-2-
\epsilon} \zeta (-3,1/2) \right. \nn \\
&  &  + \left. \epsilon (1+\epsilon)
(2+\epsilon) b^{-3- \epsilon} \zeta (-5,1/2)+ \cdots \right]
\label{lme1} \\
&+& \left( \frac{p^2}{M^2}\right)^2 B(3,3)
\left[\frac{\epsilon b^{-1-\epsilon}}{2}+
\epsilon (1+\epsilon) b^{-2-\epsilon} \zeta (-1,1/2) \right.
\nn  \\
 & & - \epsilon (1+\epsilon)(2+\epsilon) b^{-3-\epsilon}
 \zeta (-3,1/2)  \nn \\
 &  & + \left. \left.
\frac{\epsilon (1+\epsilon) (2+\epsilon)(3+\epsilon)}{2}
b^{-4-
\epsilon} \zeta (-5,1/2)+ \cdots \right] + \cdots \right\},
\nn
\eea
the numerical values of the coefficients being
\beq
B(1,1)=1, \ \ \ \ B(2,2)=\frac{1}{6}, \ \ \ \
B(3,3)=\frac{1}{30},
\ \ \ \ \frac{B(n+1,n+1)}{B(n,n)} \sim \frac{1}{4},
\eeq
and
\bea
&& \zeta (-1,1/2) =\frac{1}{24}, \ \ \ \ \ \ \zeta (-3,1/2)
=-\frac{7}{960}, \ \ \ \ \ \
\zeta (-5,1/2) =\frac{31}{8064}, \nn \\  &&  \zeta (-7,1/2) =-
\frac{127}{30720}, \ \ \ \ \ \ \zeta (-9,1/2)
=-\frac{511}{67584},
\ \ \
\ldots
\eea
We see, in fact, that the best cut of the asymp\-to\-tic ex\-pan\-si\-on
is
ob\-tain\-ed af\-ter $\zeta (-5,1/2)$ $=$ $ 0.00384$. It is apparent
also,
that the resulting regularized series will not be valid for
very small
values of $b$. On the contrary, for large values of $m_{0}/M$ the
above series is useful. In a physical setting, $m_{0}^{2} \ll
M^{2}$, so that
$b \simeq  -1/4$. In the next subsection we will obtain
convergent series for $m_{0}/M$ small (even $m_{0}=0$ is allowed) and
valid for all finite $p^{2}$.

\ms

\ni{\bf 3.2. Expansion for arbitrary momentum}.
\ms

In the general case we must follow a completely different
strategy. We shall first perform the $\epsilon$-expansion and
then
integrate over the $x$-variable in eq. (\ref{eq:integral}):
\beq
  I(\frac{p^{2}}{M^{2}}, \epsilon) = i \pi ^{2-\epsilon}
  \Gamma (\epsilon) - i \pi ^{2} J(\frac{p^{2}}{4 M^{2}}) +
  {\cal O} (\epsilon),
\eeq
where $J(z)$ is the finite part of the 1-loop integral \cite{BSh}
\beq
 J(z) = \int _{0}^{1} dx \ln [1 - 4zx(1-x)],
\eeq
which is equal to
\beq
J(z) = \left\{
\begin{array}{ll}
J_{1}(z) = 2 \sqrt{\frac{z-1}{z}} \ln (\sqrt{1-z} + \sqrt{-z}) - 2,
 & \mbox{for} \ z \leq 0, \\
J_{2}(z) = 2 \sqrt{\frac{1-z}{z}} \arctan \sqrt{\frac{z}{1-z}} - 2,
 & \mbox{for} \ 0 < z \leq 1, \\
J_{3}(z) = -i\pi \sqrt{\frac{z-1}{z}} + 2 \sqrt{\frac{z-1}{z}}
   \ln (\sqrt{z} + \sqrt{z-1}) - 2, & \mbox{for} \ z > 1. \end{array}
\right. \eeq
Now we can proceed with the summation over $l,m$.
Taking the degeneracy in (\ref{i11}) into account and using
$\zeta$-regularization for the sums, we get
\bea
\Delta I (\frac{p^{2}}{M^{2}}, \frac{m_{0}}{M},\epsilon) & = &
 i \pi^{2-\epsilon} \Gamma (\epsilon) [2 \zeta (-1,\frac{1}{2}) -1 ]
 \nonumber \\
 & - & 2 i \pi^{2} \sum_{l=1}^{\infty} (l+\frac{1}{2})
 \ln [(l+\frac{1}{2})^{2} + b] - 2 i \pi^{2} \Delta J (\frac{p^{2}}
 {4 M^{2}}; \frac{m_{0}}{M}) + {\cal O}(\epsilon).  \label{eq:sum1}
\eea
In particular, for $p^{2} < 0$
\beq
\Delta J (\frac{p^{2}}{4 M^{2}}; \frac{m_{0}}{M}) =
  \sum_{l=1}^{\infty} (l+\frac{1}{2}) J_{1} (\frac{p^{2}}{4 M_{l}^{2}}),
  \label{eq:jneg}
\eeq
while for $p^{2} > 0$
\beq
\Delta J (\frac{p^{2}}{4 M^{2}}; \frac{m_{0}}{M}) =
 \sum_{l=1}^{l^{*}(p)} (l+\frac{1}{2}) J_{3} (\frac{p^{2}}{4 M_{l}^{2}})
 + \sum_{l=l^{*}(p)+1}^{\infty} (l+\frac{1}{2}) J_{2}
 (\frac{p^{2}}{4 M_{l}^{2}})     \label{eq:jpos}
\eeq
which contains in general an imaginary part. Here $l^{*}(p)$ is the
maximum
value of $l$ which satisfies the inequality $4M^{2} (l+1/2)^{2} <
p^{2} - 4 m_{0}^{2}$. If such $l$ does not exist or is smaller than 1,
we put $l^{*}(p) = 0$ and the first sum in eq. (\ref{eq:jpos}) is
absent. As we have already mentioned, the divergent sums over $l$ are
understood as being regularized by the zeta-function regularization
procedure. These formally divergent series, which are independent
of $p^{2}$ or are linear in $p^{2}$, do not contribute to physical
(renormalized) quantities, see the discussion in Sect. 2.

The calculation is now carried out in connection with
the zeta function regularization procedure. In fact, after
expanding the functions under the summation signs in powers of
$u_{l} = p^{2}/(4 M_{l}^{2})$ we are faced up with  summations
over the
$l$-index, which give rise to Hurwitz zeta functions.
As we clearly see from expression (\ref{eq:sum1}) above, the
number of terms
contributing to each sum changes with $p$. Thus, different
explicit series are obtained for the different ranges of
$M^2/p^2$.
Specifically, for the first  ranges we get the following
series.

The first range, $|p^{2} / 4 M_{1}^{2}| < 1$, is somewhat
special and deserves a  careful
treatement. According to the preceding analysis, only
contributions in terms of a power series of $p^2/(4M^2)$ arise
in this case, and
we arrive to a series expansion which is the alternative to the
low-momentum series that was obtained before (see eq.
(\ref{lme1})),
now for small values of $m_{0}^2/M^2$, including the case
$m_{0}^2=0$, i.e.
\bea
&& \Delta I(\frac{p^2}{M^2},\frac{m_{0}}{M},\epsilon) =
 i \pi^{2-\epsilon} \Gamma(\epsilon )
 \left[ 2 \zeta (-1, \frac{1}{2}) - 1 \right]    \nn  \\
   &  & \ \ \ - 2 i \pi^{2}  \sum_{l=1}^{\infty} (l+\frac{1}{2})
   \ln \left[ (l+\frac{1}{2})^{2} + b \right]       \nn \\
 &  & \ \ \ + 4 i \pi^{2} \sum_{l=1}^\infty (l+\frac{1}{2})
 \left( \frac{u_{l}}{3} + \frac{2u_{l}^2}{15}
+ \frac{8u_{l}^3}{105} + \frac{16u_{l}^4}{315} +
\frac{128u_{l}^5}{3465}  +
\frac{256u_{l}^6}{9009} + \cdots \right) +  {\cal O}
(\epsilon) ,
 \nn \\
& & \ \ \ u_{l} \equiv \frac{p^{2}}{4 M_{l}^{2}} \equiv
\frac{\dsp \frac{p^2}{4M^2}}{(l+1/2)^2 + b}, \ \ \ \ b
\equiv \frac{m_{0}^2}{M^2} - \frac{1}{4}, \ \ \ \
  |p^2| < 4m_{0}^2 + 8M^2. \label{deltai1}
\eea
The $l$-sums yield again Epstein-Hurwitz zeta functions. In
particular,
\beq
 2 \sum_{l=1}^{\infty} (l + 1/2)   u_{l}^k = \frac{1}{2(1-
k)} \left. \frac{\p}{\p a} E_1^{(1)} (k-1;a, b) \right|_{a=1/2} \,
\left( \frac{p^2}{4M^2}\right)^k,
\eeq
where the superindices $(1)$ mean `truncated', in the sense
that the first term in the definitions of these zeta functions
(the one for $n=0$) is absent, namely
\beq
E_1^{(1)} (k; a, b) = \sum_{n=1}^\infty [(n+a)^2+b]^{-k}
=F(k+1; a,b),
\eeq
in terms of the function $F$ introduced before. Let us call
\beq
  h^{(1)}(k;1/2,b) \equiv \left. \frac{1}{1-k} \frac{\p}{\p a}
  E^{(1)}_{1} (k;a,b) \right|_{a=1/2}.
\eeq
The following bounds for these coefficients of
the power series expansion will be useful. First,
introducing the constants
\beq
\alpha_k \equiv  \sum_{l=1}^{\infty} (2l+1) [l (l+1)]^{-k} <
    2 \zeta (2k-1),
\eeq
we have
\bea
&& \alpha_1 =1.1544, \ \ \alpha_2 =0.9996, \ \  \alpha_3
=0.4041,
\ \ \alpha_4 =0.1918, \ \  \alpha_5 =0.0944,  \nn \\
&& \alpha_6 =0.0470, \ \ \alpha_7 =0.0235, \ \  \alpha_8
=0.0117, \ \ \alpha_9 =0.0059, \ \ldots,
\eea
and
\bea
&& m_{0}=0, \ \ M\neq 0 \ \mbox{arbitrary}:
\ \ \ \ h^{(1)}(k;1/2,b) = \alpha_k, \nn \\
&& m_{0} \neq 0 \ \ \left\{ \begin{array}{ll} m_{0}^2 \leq M^2/4: & \ \
2 \zeta^{(1)} (2k-1, 1/2) \leq  h^{(1)}(k; 1/2, b) \leq
\alpha_k,
\\  m_{0}^2 \geq M^2/4: & \ \  h^{(1)}(k; 1/2, b) \leq 2\zeta^{(1)}
(2k, 1/2). \end{array} \right.
\eea
We obtain, in particular
\bea
h^{(1)}(1; 1/2, b) |_{b \geq 0} &\leq & 0.93 \leq
h^{(1)}(1;1/2,b)|_{-1/4 \leq b \leq 0} \leq 1.15, \nn \\
h^{(1)}(2; 1/2, b) |_{b \geq 0} &\leq & 0.83 \leq
h^{(1)}(2;1/2,b)|_{-1/4 \leq b \leq 0} \leq 0.99, \nn \\
h^{(1)}(3; 1/2, b) |_{b \geq 0} &\leq & 0.29 \leq
h^{(1)}(3;1/2,b)|_{-1/4 \leq b \leq 0} \leq 0.40, \nn \\
h^{(1)}(4; 1/2, b) |_{b \geq 0} &\leq & 0.12 \leq
h^{(1)}(4;1/2,b)|_{-1/4 \leq b \leq 0} \leq 0.19, \nn
\eea
In few words, we see that with increasing $m_{0}$ the value of the
coefficients decreases, starting from very reasonable values
(the $\alpha_k$) for $m_{0}=0$. However, the fact that $b$ in eq.
(\ref{deltai1})
can be negative (for small values of $m_{0}$) makes it difficult
in practice
to use expression (\ref{eq:repres}) for the Epstein-Hurwitz zeta
function. A more
convenient (albeit lengthly) alternative is to perform a
binomial expansion, which is absolutely convergent in
$p^{2}/4 M_{1}^{2}$ whenever $|b(2/3)^{2}| < 1$,
i.e. $m_{0}^{2}/ M^{2} < 5/2$ \cite{Casimir}, \cite{camporesi},
\cite{eeb3}, \cite{got}. In this way, we
obtain \bea
 \Delta I (\frac{p^2}{M^2},\frac{m_{0}}{M},\epsilon) &=&
 2 i \pi^2 \left[ - \frac{11}{24} \pi^{-\epsilon} \Gamma(\epsilon )
+ 2 \zeta' (-1, 1/2) + \ln 2
\right. \nn \\
&+& \sum_{k=1}^\infty \frac{(-1)^{k}}{k} \,
\zeta^{(1)} (2k-1, 1/2) \left(
\frac{m^2_0}{M^2} - \frac{1}{4}
\right)^k \nn \\
  &+& \left. \sum_{k=1}^\infty c_k \left( \frac{m_{0}^{2}}{2 M^{2}} + 1
  \right)^{k} h^{(1)}
\left(k;\frac{1}{2}, \frac{m^2_0}{M^2} - \frac{1}{4}\right)
\left( \frac{p^2}{4M_{1}^2}\right)^k
+ { O} (\epsilon^2 ) \right],
\label{1-3}
\eea
where
\bea
&& c_1 = \frac{2}{3}, \ \ c_2 = \frac{2^{3}}{3 \cdot 5}, \ \
 c_3 = \frac{2^6}{3\cdot 5 \cdot 7}, \ \ c_4 = \frac{2^8}{5 \cdot 7
\cdot 9}, \nn \\
&& c_5 = \frac{2^{12}}{5\cdot 7 \cdot 9 \cdot 11}, \ \ c_6 = \frac{2^{20}}
{7\cdot 9 \cdot 11 \cdot 13}, \cdots
\eea

For higher positive values of $p^2$ contributions of new type appear
(corresponding to the first sum of eq. (\ref{eq:jpos})). Thus, for
$4 (m_{0}^{2} + 2M^{2}) \leq p^{2} < 4 (m_{0}^{2} + 6 M^{2})$ we get
\bea
 \Delta I \left( \frac{p^2}{M^2},\frac{m_{0}}{M},\epsilon \right) & = &
  i \pi^{2-\epsilon} \Gamma (\epsilon )
    \left[ 2 \zeta(-1,\frac{1}{2}) - 1  \right]
     -  2i \pi^{2} \sum_{l=1}^{\infty} (l+\frac{1}{2})
    \ln [(l+\frac{1}{2})^{2}+b]  \nn \\
    & - & i \pi^{2} \left[
 i \pi \sqrt{1-\frac{4M_{1}^{2}}{p^{2}}}
    + \sqrt{1-\frac{4M_{1}^{2}}{p^{2}}} \ln \frac{p^{2}}{M_{1}^{2}}
                                    \right. \nn \\
    & - & \left. \frac{2M_{1}^{2}}{p^{2}} \left( 1 -
    \frac{M_{1}^{2}}{2p^{2}} + \ldots \right)-2
    + 2\sum_{l=2}^{\infty} (l+\frac{1}{2}) J_{2}
    (\frac{p^{2}}
    {4 M_{l}^{2}}) \right] + {\cal O} (\epsilon) \nn \\
    & = &
 2 i \pi^2 \left[ - \frac{11}{24} \pi^{-\epsilon} \Gamma(\epsilon )
+ 2 \zeta' (-1, 1/2) + \ln 2
\right. \nn \\
&+& \sum_{k=1}^\infty \frac{(-1)^{k}}{k} \,
\zeta^{(1)} (2k-1, 1/2) \left(
\frac{m^2_0}{M^2} - \frac{1}{4}
\right)^k - \frac{3}{2} \sqrt{1-\frac{4M_{1}^{2}}{p^{2}}}   \\
  &+& \frac{1}{2}\sqrt{1-\frac{4M_{1}^{2}}{p^{2}}} \ln
       \frac{p^{2}}{M_{1}^{2}}
    - \frac{M_{1}^{2}}{p^{2}} \left( 1 - \frac{M_{1}^{2}}{2p^{2}}
    + \ldots \right)  \nn \\
  &+& \left. \sum_{k=1}^\infty 3^{k} c_k
  \left( \frac{m_{0}^{2}}{6 M^{2}} + 1 \right)^{k}
  h^{(2)}
\left(k;\frac{1}{2}, \frac{m^2_0}{M^2} - \frac{1}{4}\right)
\left( \frac{p^2}{4M_{2}^2}\right)^k
+ { O} (\epsilon^2 ) \right],   \nn
\eea
and so on, for the rest of the intervals for higher momentum.
Here
\beq
  h^{(2)}(k;1/2,b) \equiv \left. \frac{1}{1-k} \frac{\p}{\p a}
  E^{(2)}_{1} (k;a,b) \right|_{a=1/2},
\eeq
being $E_1^{(2)}$ the truncated function
\beq
E_1^{(2)} (k; a, b) \equiv \sum_{n=2}^\infty [(n+a)^2+b]^{-k}.
\eeq
Alternatively, the whole expression can be expanded in terms
of the
 truncated Hurwitz zeta functions
\beq
 \zeta^{(k)} (s, 1/2) \equiv  \zeta (s, 1/2) -
\sum_{n=0}^{k-1} (n+
1/2)^{-s}, \ \ \ \ k=1,2,3, \ldots
\eeq
whose numerical values are:
\bea
&&   \zeta^{(1)} (2, 1/2) \simeq 0.93, \ \ \  \zeta^{(1)} (4,
1/2)
\simeq 0.23, \  \ \  \zeta^{(1)} (6, 1/2) \simeq 0.09, \nn  \\
&& \zeta^{(1)} (8, 1/2) \simeq 0.04, \ \ \
   \zeta^{(2)} (2, 1/2) \simeq 0.4904,  \ \ \
   \zeta^{(2)} (4,1/2) \simeq 0.0373,  \nn \\
&&  \zeta^{(2)} (6, 1/2) \simeq 0.0048,  \ \ \
  \zeta^{(2)} (8, 1/2) \simeq 0.0007,  \ \ \
  \zeta^{(3)} (2, 1/2) \simeq 0.33036, \nn \\
&& \zeta^{(3)} (4,1/2) \simeq 0.01172,  \ \ \
   \zeta^{(3)} (6, 1/2) \simeq 0.00073, \ \ \
  \zeta^{(3)} (8, 1/2) \simeq 0.00005. \nn
\eea
Furthermore, a useful asymptotic expression for the
derivative
$\zeta' (-1,1/2)$ can be found in \cite{eli2}.
By analogous methods, expansions for $p^{2} < 0$ can be obtained.
\bs

\section{Calculation of the total cross section}

In this section we calculate the total cross section
$\sigma^{(\infty)}(s)$ of the scattering process \
(2 {\em light particles}) \ $\longrightarrow $ \
(2 {\em light particles}), \  in the case when
the whole Kaluza-Klein tower of heavy particles contribute, and we
compare it with $\sigma ^{(N)} (s)$, the cross section obtained
for the case when only $N$ first modes are taken into account (i.e.
modes with $l=0,1,2, \dots , N$). With this notation, $\sigma ^{(0)}
(s)$ is the cross section in the dimensionally reduced model, i.e.
when only the light particle contributes. Such comparison will
be quite illuminative for understanding the relative contributions
of the various heavy modes.

We have found that the quantity which describes the net effect due
to the
tower of heavy particles is the following ratio, which is built up of
the total cross sections:
\beq
 \Delta^{(\infty,0)} \left(\frac{s}{4M^{2}};\frac{\mu_{s}^{2}}{M^{2}},
 \frac{\mu_{u}^{2}}{M^{2}},\frac{\mu_{t}^{2}}{M^{2}};
 \frac{m_{0}}{M} \right) \equiv 16 \pi^{2} \frac{\sigma^{(\infty)}(s) -
 \sigma^{(0)}(s)}{ g  \sigma^{(0)}(s)} .   \label{eq:delta-def}
\eeq
Using the expression for the
4-point Green function (\ref{eq:gfren2}), renormalized
according to (\ref{eq:renorm1}) and (\ref{eq:renorm2}), we
calculate the corresponding total cross sections and obtain that,
to leading order (i.e. 1-loop order)
in the coupling constant $g$, the function (\ref{eq:delta-def}) is equal to
\bea
 && \Delta^{(\infty,0)}
\left(\frac{s}{4M^{2}};\frac{\mu_{s}^{2}}{M^{2}},
 \frac{\mu_{u}^{2}}{M^{2}},\frac{\mu_{t}^{2}}{M^{2}};
 \frac{m_{0}}{M} \right)   =
  - 2 \left\{ Re \Delta J (\frac{s}{4 M^{2}}, \frac{m_{0}}{M})  \right.\nn \\
 \ \ \ \ &  & + \frac{2}{s-4m_{0}^{2}} \int_{-(s-4
m_{0}^{2})}^{0} du  \Delta J (\frac{u}{4M^{2}};\frac{m_{0}}{M})
   - \Delta J (\frac{\mu_{s}^{2}}{4M^{2}}; \frac{m_{0}}{M}) \nn \\
 \ \ \ \ &&  - \left. \Delta J (\frac{\mu_{t}^{2}}{4M^{2}};
\frac{m_{0}}{M})
   - \Delta J (\frac{\mu_{u}^{2}}{4M^{2}}; \frac{m_{0}}{M})
   \right\}.  \label{eq:delta-exp}
\eea
Here we assume that $\mu_{s}^{2},\mu_{t}^{2},\mu_{u}^{2} < 4 m_{0}^{2}$.

For the numerical evaluation we take the zero mode particle to be much
lighter than the first heavy mode and choose the subtraction point to be
at the low energy interval, even in comparison with $m_{0}$. Recall that
$\mu_{s}^{2}+\mu_{t}^{2}+\mu_{u}^{2} = 4m_{0}^{2}$. We take
\beq
   \frac{m_{0}^{2}}{M^{2}} = 10^{-4}, \  \  \
   \frac{\mu_{s}^{2}}{m_{0}^{2}} = 10^{-2}, \  \  \
\mu_{u}^{2}=\mu_{t}^{2}=\frac{1}{2} (4 m_{0}^{2} - \mu_{s}^{2}).
\label{eq:param}
\eeq

First of all, we observe that if the parameters of our theory
satisfy $\mu_{s}^{2} \ll m_{0}^{2} \ll M^{2}$ (notice that for our
choice
(\ref{eq:param}) these inequalities are indeed fulfilled) the dependence
of the function $\Delta ^{(\infty,0)}$ on $\mu_{s}^{2}/M^{2}$,
$\mu_{t}^{2}/M^{2}$, $\mu_{u}^{2}/M^{2}$ and $m_{0}^{2}/M^{2}$ is very
weak, and in practice it depends only on one dimensionless parameter.
We choose this parameter to be $z=s/(4 M_{1}^{2})$.

Using the results of Sect. 3 we calculate the function
$\Delta ^{(\infty,0)}(z)$. Its plot in the range $0 < z < 1$ is
presented in Fig. 1. We see that the contribution of the
Kaluza-Klein tower of particles
is considerable.  Thus, $\Delta ^{(\infty,0)} \simeq  0.51$
for $s = 0.5 (4 M_{1}^{2})$ and $\Delta ^{(\infty,0)} \simeq  0.12$
for $s = 0.25 (4 M_{1}^{2})$. So, already for the energies much smaller
than the threshold of the first heavy particle the effect is quite
noticeable.

To be remarked is also the quick convergence of the
sums over $l$ in (\ref{eq:jneg}) and (\ref{eq:jpos}), contributing to
the function $\Delta^{(\infty,0)}(z)$.
For instance, a few terms
give already curves which do not change any more when adding
more summands (we have checked indeed that the sum of 50 terms of the
above series is, for all purposes, identical to the sum of the 20 first
terms).

Of course, due to the convergence of the sums only the first few terms
with low $l$ give essential contributions whereas those corresponding to
higher modes are quite negligible. To understand how many modes
we really see from the plot of $\Delta ^{(\infty,0)}$ further
analysis is needed. For this purpose, in Fig. 3 we also present
the curve $\Delta ^{(1,0)}(z)$ characterizing the contribution
of the first heavy mode only. We see that the difference is not
small ($0.3$ for $z=0.5$ and $0.05$ for $z=0.25$), and the function
$\Delta ^{(\infty,0)}(z)$ {\it within a few per cent accuracy}
represents {\it more} than just the first mode.

To have more illustrative characteristics, we introduce the
quantities
\beq
   \epsilon _{N} (z) \equiv \frac{\Delta ^{(N,0)}(z)}
   {\Delta ^{(\infty,0)}(z)},     \label{eq:epsilon-def}
\eeq
which show the relative contributions of the first $N$ heavy
Kaluza-Klein modes. The plots for some $\epsilon _{N}(z)$
for $0<z<1$ are presented in Fig. 4. We conclude that with
an  accuracy of about $5 \div 10 \% $, the function
$\Delta ^{(\infty,0)}(z)$ in the range
$s \sim 0.8 M_{1}^{2} \div 2.4 M_{1}^{2}$ actually
shows the presence of at least $3 \div 4$ first
heavy modes in the theory.

Another interesting question is how the contribution
of the heavy Kaluza-Klein modes depends on the topology
of the space of extra dimensions. Here we restrict
ourselves only to comparison of the results
for $\Delta _{S}^{(\infty,0)}(z)$, which is calculated in this
article (``$S$" stands for
``spherical compactification''), with the behaviour of $\Delta
_{T}^{(\infty,0)} (z)$ for the toroidal compactification calculated in
ref. \cite{DKPC} for the similar model but with the space-time being
$M^{4} \times T^{2}$, where $T^{2}$ is the two-dimensional torus.
It is clear that the results of the comparison depend on the relation
between the inverse radius of the sphere $M_{S} = L^{-1}$ (for further
discussion we have
attached the index "$S$" to it) and the scale $M_{T}$ equal to the
inverse radius of the circles forming the torus $T^{2} = S^{1} \times
S^{1}$. For the present analysis we assume that in both compactifications
the dimensionally reduced models, i.e. zero mode sectors of the
initial theories, coincide and that the masses of the first heavy modes
are the same. These imply that for both cases the mass $m_{0}$
of the zero mode is the same and
\beq
   2 M_{S}^{2} = M_{T}^{2}.     \label{eq:S-T-scales}
\eeq
The plots of the functions $\Delta_{S}^{(\infty,0)}$ and
$\Delta_{T}^{(\infty,0)}$ for $0 < z < 1$ are represented in Fig. 5.
Table 1 provides explicit values of these functions, corresponding to a
sample of values of $z$.
The difference between the curves is quite noticeable (for example,
$\Delta_{T}^{(\infty,0)} - \Delta_{S}^{(\infty,0)} = 0.25$ for $z=0.5$
and is equal to $0.06$ for $z=0.25$) and is due to the difference
of the spectra of the Laplace operator on the sphere and on the torus.
Namely, for the two-dimensional sphere $S^{2}$ the eigenvalues $\lambda_{l}
(S^{2})$ of the Laplace operator, the squares of the
masses of the Kaluza-Klein modes determined by them, and the
multiplicities $d_{l}(S^{2})$
of the eigenvalues are given by (cf. eq. (\ref{eq:mass}))
\beq
   (M_{l}^{(S)})^{2} = m_{0}^{2} + \lambda_{l}(S^{2}) M_{S}^{2},  \; \; \;
   \lambda_{l}(S^{2}) = l(l+1), \; \; \;
   d_{l}^{(S)} = 2l+1 \ .
\eeq
For the two-dimensional torus $T^{2}$ the analogous are:
\beq
   (M_{n}^{(T)})^{2} = m_{0}^{2} + \lambda_{n}(T^{2}) M_{T}^{2}, \; \; \;
   \lambda_{n}(T^{2}) = n_{1}^{2}+n_{2}^{2},
\eeq
where $n=(n_{1},n_{2})$ is a two-vector labelling the eigenvalues,
$-\infty < n_{i} < \infty$ $(i=1,2)$, and
$d_{|n|}^{(T)}$ is the number of such vectors with the same length
$|n|=\sqrt{n_{1}^{2}+n_{2}^{2}}$. Using the expansions similar to
(\ref{deltai1}) and eq. (\ref{eq:delta-exp}) we get that for $|z|<1$
\beq
\Delta_{K}^{(\infty,0)}(z) \approx \frac{4}{9} \left(z \frac{M_{T}^{2}}
 {M_{K}^{2}}\right)^{2} \zeta (2|K).    \label{eq:asymp}
\eeq
Here the index $K$ labels the type of compactification, e.g. $K=S$
for the case of the sphere and $K=T$ for the case of the torus, and
$\zeta(s|K)$ is the zeta-function of the Laplace operator on the
manifold $K$ \cite{hawk} (see also \cite{zeta-fun})
\beq
   \zeta(s|K) = {\sum_{n}}' \frac{1}{(\lambda_{n}(K))^{s}},
\eeq
where the prime means that the term corresponding to the zero eigenvalue
is absent from the summation. For example, for $K=S$ this function
can be expressed in terms of the derivative of the generalized
Epstein-Hurwitz zeta-function (\ref{fm}):
\bea
  \zeta (s| S^{2}) & = & \sum_{l=1}^{\infty} \frac{ d_{l}(S^{2})}
     { \left[ \lambda _{l}(S^{2}) \right]^{s}} =
     \sum_{l=1}^{\infty} \frac{2l+1}{[l(l+1)]^{s}}   \nn  \\
     & = & - \frac{1}{s-1} \frac{\p}{\p a}
     F(s;a,-\frac{1}{4}) |_{a=1/2} \ .   \label{eq:S-T-ratio}
\eea
Taking into account the relation (\ref{eq:S-T-scales}) between
$M_{S}^{2}$ and $M_{T}^{2}$, we obtain an approximate expression
relating the ratio of the contributions of the Kaluza-Klein towers of
particles corresponding to the spherical and toroidal
compactifications with the
characteristics of the Laplace operator on these manifolds:
\beq
  \frac{\Delta_{S}^{(\infty,0)}(z)}{\Delta_{T}^{(\infty,0)}(z)} \approx
  \frac{4 \zeta(2|S^{2})}{\zeta(2|T^{2})} \approx 0.66 \ .
\eeq
Results of our numerical computations (Fig. 5) are with good
accuracy in accordance with formula (\ref{eq:S-T-ratio}).

In spite of the fact that, from the point of view of high-energy
physics, it seems rather unlikely
that  energies satisfying $\sqrt{s} > 2M_{1}$ will be available
in the nearest future, for theoretical
reasons and for the sake of completeness we
have also calculated the function $\Delta_{S}^{(\infty,0)}(z)$ for
$1<z<21$ (see Fig. 6). Peaks of the curve correspond very
approximately
(since $m_{0} \neq 0$) to the thresholds of creation of heavy mode
particles, i.e. to the values $s = 4 M_{l}^{2}$ or $z=l(l+1)/2$ for
$l=1,2,3,4,5,6$.
\begin{table}

\begin{center}

\begin{tabular}{|c||c|c|c|c|c|}
\hline \hline
 $z$  & 0.1 & 0.25 & 0.5 & 0.75 & 0.99 \\
 \hline\hline
$\Delta_S^{(\infty, 0)}$ & 0.0181 & 0.1165 & 0.5111 & 1.361 & 3.969 \\
 \hline
$\Delta_S^{(1, 0)}$ & 0.0136 & 0.0888 & 0.3982 & 1.103 & 3.508 \\
 \hline
$\Delta_T^{(\infty, 0)}$ & 0.0271 & 0.1749 & 0.7626 & 2.008 & 5.650 \\
\hline \hline \end{tabular}

 \caption{{\protect\small Table of numerical values.}}

\end{center}

\end{table}

\section{Conclusions}

In this paper we have studied the behaviour of the total
cross section for scattering of two light particles in an
effective theory in four dimensions, obtained from the
six-dimensional scalar theory through the spherical compactification
of two extra dimensions.

Though our model cannot be directly termed as being physical, we
do believe that the
effect we have calculated is of a very general nature, and  that it will
also take place
in more realistic theories. Thus, our results can be {\em in
principle} used for comparison with actual experiments. The idea is the
following.

We assume that the low-energy sector of the theory
is already well determined. Therefore,
the value of the renormalized coupling constant $g$ is known
and the total cross
section $\sigma ^{(0)} (s)$ can
be calculated with sufficient accuracy.
Experimentally one
should measure the total cross section $\sigma ^{exp}(s)$ and compute
the quantity
\[
\Delta ^{exp}(s)= - 16 \pi^{2} \ \frac{\sigma^{exp}(s)-
\sigma^{(0)}(s)}{g \sigma^{(0)}(s)} \ .
\]
(cf. (\ref{eq:delta-def})). If above the threshold for the light
particle, one has that $\Delta ^{exp} (s) = 0$, then
there is no evidence of heavy Kaluza-Klein modes at given
energies.

If, on the contrary, $\Delta ^{exp} (s) \neq 0$, then there will be
evidence for the existence
of heavier particles. The obvious next step would be to see which curve
$\Delta _{K}^{(N,0)}(s)$ fits the
experimental data best. If it is the curve with $N=\infty$ (or
sufficiently large $N$), for a certain manifold $K$, this fact
should be considered as an indirect
evidence of the {\em multidimensional} nature of the interactions ---at
least within the framework of the given class of models and
for that type of compactification.

Our calculations suggest that, indeed, the effect can be quite
noticeable, even for energies
below the threshold of the first heavy particle (see Fig. 1). The values
of the parameters (\ref{eq:param}) that we have used for our
computations can
mimic a physical situation with, for example, $m_{0}=100$ GeV and $M=
10$ TeV, and with the charge $g$ renormalized at the low energy point
$\sqrt{\mu_{s}^{2}} = 10$ GeV.

Our results also show that we can distinguish between
different types of compactification of the extra dimensions. The main
contribution, for energies below the threshold of the first heavy
mass state, is basically determined by the zeta-function $\zeta(2|K)$,
uniquely associated to the two-dimensional manifold $K$ through the
spectrum of the Laplace operator on it. This provides, by the way, a
further example of the relevance of the concept of zeta
function in high-energy physics.

Of course, one should not forget that the effect studied here is
of one-loop order and, apparently, rather hard to detect
experimentally.
 Because of this, an interesting possibility in a more realistic
model would be to consider specific processes for which the tree
approximation is absent and the leading contribution is given by
the one-loop diagrams even in the zero-mode sector of the theory.
Such processes are obviously more sensitive to the heavy Kaluza-Klein
modes.

\vspace{5mm}

\ni{\large \bf Acknowledgments}

It is a pleasure to thank F. Boudjema, A. Demichev, K. Kirsten,
C. Nash and
D. O'Connor for interesting discussions on the problems
addressed in this paper.
Yu.K. thanks the Department ECM of Barcelona University for warm
hospitality.
This investigation has been supported by DGICYT (Spain), project No.
PB90-0022 and sabbatical grant
SAB 92 0267, and by CIRIT (Generalitat de Catalunya).
\bs

\newpage

\section*{Figure captions}

\begin{description}
  \item[Fig. 1] Tree diagrams contributing to the 4-point Green function
                $\Gamma ^{(\infty)}$. The lines correspond to the light
                particle, the bar corresponds to derivatives
                with respect to external momenta.
  \item[Fig. 2] 1-loop diagrams contributing to the 4-point Green
                function $\Gamma ^{(\infty)}$. Thin lines correspond
                to the light particle with mass $m_{0}$. The thick line
                with the label $M_{l}$ corresponds to propagation
		of the particle with mass $M_{l}$.
  \item[Fig. 3] Plots of the functions $\Delta ^{(\infty,0)}(z)$ and
                $\Delta ^{(1,0)}(z)$ in the interval $0<z<1$ for the
                sphere compactification.
  \item[Fig. 4] Plots of the functions $\epsilon_{N}(z)$ defined by
                eq. (\ref{eq:epsilon-def}) for $N=1,2,3,4,5$;
                $\epsilon_{\infty}(z) \equiv 1$.
  \item[Fig. 5] Plots of the functions $\Delta_{S} ^{(\infty,0)}(z)$ and
                $\Delta_{T} ^{(\infty,0)}(z)$ in the interval
                $0<z<1$ for compactifications of extra dimensions
		to the two-dimensional sphere $S^{2}$ and the
		two-dimensional torus $T^{2}$ respectively. In
		both cases the dimensionally reduced models are the
		same and scales characterizing these manifolds
		are chosen in such a way that $2 M_{S}^{2} =
		M_{T}^{2}$, so that the masses of the first heavy
		particles of the Kaluza-Klein towers are equal.
  \item[Fig. 6] Plot of the function $\Delta_{S}^{(\infty,0)}(z)$ with
                the sphere compactification for $0<z<21$, i.e. up to
                the threshold of the sixth heavy mode.
\end{description}

\end{document}